\documentstyle[preprint,aps]{revtex}
\begin{document}

\draft

\title{Polarized interacting exciton gas in quantum wells and
bulk semiconductors}

\author{J.Fern\'{a}ndez-Rossier and C.Tejedor}

\address{Departamento de F\'{\i}sica de la Materia Condensada,
Universidad Aut\'onoma de Madrid, Cantoblanco, 28049, Madrid, Spain.}

\author{L.Mu\~noz and L.Vi\~na}

\address{Departamento de F\'{\i}sica de Materiales,
Universidad Aut\'onoma de Madrid, Cantoblanco, 28049, Madrid, Spain.}

\date{\today}

\maketitle

\begin{abstract}

We develop a theory to calculate exciton binding energies of both two- and
three-dimensional spin polarized exciton gases within a mean field approach.
Our method allows the analysis of recent experiments showing the importance
of the polarization and intensity of the excitation light on the exciton
luminescence of GaAs quantum wells. We study the breaking of the spin
degeneracy
observed at high exciton density $(5 \ \ 10^{10} cm ^2)$.
Energy level splitting betwen spin +1 and spin -1
is shown to be due to many-body inter-excitonic exchange while the
spin relaxation time is controlled by intra-exciton exchange.

\end{abstract}

\section{Introduction}

The optical response of intrinsic semiconductors heterostructures has
received considerable attention in recent years from both the
theoretical and the experimental point of view. The study of the
luminiscence spectrum gives information about the lowest electronic
excited state of a semiconductor, i.e. the exciton.
Using polarized light, emited photons contain information on both exciton
energies and their dependence on the exciton spin.\cite{coms}
Time resolved photoluminescence experiments \cite{1LV,2LV,3Aman,4Aman,Stark}
provide information on different exciton
properties: exciton formation and decay processes, spin relaxation,
binding energy evolution, etc.  In some of those experiments,
performed in the picosecond range,\cite{1LV,2LV,3Aman,4Aman,Stark}
an energy splitting between spin +1 and spin -1 excitons has been reported.
These studies show that the spin splitting increases with both the initial
exciton density and the degree of initial polarization.

So far, theoretical models have been proposed\cite{7Sh,BAP} to explain only
the exciton spin relaxation without taking into account the exciton-exciton
interaction. The models give alternative explanations to spin relaxation
processes in terms of intraexcitonic exchange,\cite{DM,Knox,PB,KZ}
D'yakonov-Perel\cite{DP} and Elliot-Yaffet\cite{EY} mechanisms.
However, these free-exciton models fail to describe the spin level splitting.

As far as many-exciton effects are concerned, only
the {\em spinless} high density exciton gas has been the subject of
theoretical research in the last 30 years.\cite{5Ha,9No,6Sc,8Ke}
Three different schemes have been developed.
The first one, proposed by Keldysh\cite{8Ke} and generalized by
Comte and Nozieres,\cite{9No} is a BCS like approach.
Another method, due to Hanamura and Haug,\cite{5Ha}
consists in writing the Hamiltonian in terms of
exciton operators. When the center of mass momentum of
each exciton is zero these two approaches
are equivalent up to order $(na^{d})^{2}$ where $n$ is the exciton densty,
$a$ is the exciton radius $\ (\hbar)^2 \epsilon /\mu e^2$, $e$ being
the electron charge, $\mu$  the reduced exciton mass and $\epsilon$
the dielectric constant, and $d$ is the dimension of the space. The last
availaible theoretical scheme consists in  writing a Bethe-Salpeter
equation and interpreting the homogeneous part as a multi-exciton Wannier
equation.\cite{10Zi,13Kr} The physics
underlying the three approaches is always a mean field treatment
of interaction between {\em spinless } excitons and so the equations
obtained are analogous. The differences lie in the obtention of the equations
and in the physical nature of the mathematical objetcs the theories are built
up.
In any case, spin splitting is beyond the scope of those {\em spinless}
excitons theories.

We present in this paper a theory of spin-dependent exciton-exciton
interaction in two and three dimensions (2D and 3D).
Such interaction produces in a gas
with a difference in the spin populations, a level splitting.
We concentrate on the spin and energy dynamics and
their effect on the exciton recombination without paying attention to the
exciton formation. Our theory for spin polarized systems is an extension of the
exciton operator Hamiltonian approach (Hanamura-Haug approach\cite{5Ha}).
Our results for the 2D case are in agreement with
experiments in quantum wells.\cite{1LV,2LV,3Aman,4Aman,Stark}
Experimental work in the interacting regime with polarized light,
remains to be done in bulk, to the best of our knowledge.

This paper is organized as follows. In section II, we present the
many-exciton Hamiltonian\cite{6Sc} from  which we obtain many-exciton
Wannier equations including  spin. In section III we solve them
using both perturbative and variational approaches.
The main approximation we make is to consider the exciton center of mass
at rest.  Our calculations, correct up to order $(na^{d})^{2}$,
give four exciton energies and wave functions as a function of
density in 2 and 3 dimensions. The maximum density experimentally
reached is about $10 ^{11}$ excitons $cm ^{-2}$.
A reasonable estimate for the exciton radius is $100 \AA$.
When $n =10 ^{11}$ $cm^{-2}$ then $ na ^{d}$ is roughly 0.1.
This means we neglect $10^{-2}$ compared with $10^{-1}$.
Screening corrections to the energy within RPA are
also taken into account altough they do not depend on the exciton spin.
A rate equation is proposed and solved in section IV in order to obtain
the time evolution of the different types of excitons.
Once we have the energies as a function of densities and the densities as a
function of time, we can get results directly comparable with experiments
obtaining a qualitatively good agreement. Finally, section V is devoted
to the discussion and summary of our main results.

\section{Interacting Polarized Exciton Gas Theory }

Dealing with exciton-exciton interaction is a complicated task. The exciton is
a two-body excitation which is not, properly speaking, a boson.
Even the non interacting
exciton Green function is not simple to calculate.\cite{15Ri,16Mac,14Mah}
Consider the exciton creation operator:
\begin{equation}
\psi^{\dagger} _{i} =\int de\ dh\ \phi _{i}(e,h) \psi^{\dagger} _{h}
\psi^{\dagger}_{e}
\end{equation}
where the index $i$ is the set of exciton quantum
numbers $\{\ {\bf K} ,J, M,\nu \}$, ${\bf K}$ being the center of
mass momentum, $J$ the total angular momentum, $M$
its third component and $\nu $ labeling the internal state of the exciton.
$\psi^{\dagger}_{e}$ creates
a conduction band electron in $e$, where $ e=\{\ {\bf r}, s _{e} \}\ $,
and analagously $\psi^{\dagger}_{h}$ for a hole for which we neglect
valence band mixing effects. On the
other hand, $\phi _{i}(e,h)$ is the exciton wave function.
So, physically speaking, $ \psi^{\dagger} _{i}$ is an
operator that creates a electron-hole cloud following the excitonic
probabilty amplitude. The exciton wave function can be written
as a product of  factors, namely, the center of mass term, the orbital part
and the radial wave function. This factorization is possible if we
assume that $\{\ {\bf K}, J, M, \nu \}$ are good quantum
numbers for the non-interacting exciton.

Now it can be easily checked that
exciton operators fulfill the following commutation relation \cite{6Sc}
\begin{eqnarray}
\left[ \psi _{i} , \psi _{i'} \right] = 0 ;
\  \  \left[ \psi ^{\dagger } _{i} , \psi ^{\dagger } _{i'} \right] = 0
\nonumber \\
\left[ \psi _{i} , \psi ^{\dagger}_{i'} \right] = \delta _{i,i'}- \int
de\ dh\ \phi ^{*}_{i}(e,h) \{ \int dh' \phi _{i'}(e,h') \psi ^{\dagger}_{h}
\psi _{h} \ + \int de' \phi _{i'}(e',h) \psi ^{\dagger}_{e} \psi _{e} \}
\label{CR}
\end{eqnarray}
The second term of the left hand side of eq. \ref{CR} is an operator
whose expectation value in a state with $n$ excitons
is of the order $na ^{d}$. In the limit of vanishing density, the
excitons are strictly bosons. Up to order $(na ^{d}) ^{2}$
it can be shown \cite{6Sc} that  the exciton operator Hamiltonian including
the exciton exciton interaction can be written as:

\begin{eqnarray}
H=H _0 +H _{int}
\label{Hamil}
\end{eqnarray}
where
\begin{eqnarray}
H _{0}=\sum _{i,i'} <i \mid T _{e} +\ T _{h} -\ V _{eh}| i'> \psi
^{\dagger}_i \psi _{i'}
\end{eqnarray}
and
\begin{eqnarray}
H _{int}=  \frac{1}{2} \sum_{i,i',i'',i'''} \psi ^{\dagger} _i
\psi ^{\dagger} _{i'} \psi _{i''} \psi _{i'''} (<i,i'|I _d | i'',i'''> +
<i,i' | I _x |i'',i'''> )
\end{eqnarray}
where
\begin {eqnarray}
<i |T _{e} + T _{h} - V _{eh}|i'> = \int de dh   \phi ^{*}_{i}(e,h) \ \
(T _{e} + T _{h} - V _{eh}  ) \phi _{i'}(e,h)
\nonumber \\
<i,i'|I_d|i'',i'''>= \int de dh de' dh' \phi ^{*}_{i}(e,h) \phi
^{*}_{i'}(e',h')(V_{ee'}+V_{hh'}-V_{eh'}-V_{he'}) \phi_{i''}(e,h) \phi _{i
'''}(e',h')\nonumber \\
<i,i' | I_x | i'',i'''> = \! - \! \int \! de dh de' dh' \ \phi
^{*}_{i}(e,h) \phi
^{*}_{i'}(e',h')(V_{ee'}+V_{hh'}-V_{eh'}-V_{he'}) \phi_{i''}(e',h) \phi
_{i'''}(e,h')
\end{eqnarray}
It must be stressed that this Hamiltonian is correct up
to order $(na ^{d}) ^{2}$. This means that using this
Hamiltonian we can (and we must) neglect the contributions to the energy
of order $(na ^{d})^{2}$. The physical origin of $I_d$ is the direct
unscreened Coulomb interaction between fermions belonging to different
excitons while $I_x$ is the inter-excitonic exchange, or the unscreened
exchange interaction between fermions of the same type. It is important to
distingish between inter-excitonic exchange and intra-excitonic exchange.
The former is, as we stated before, a many exciton intraband
(conduction-conduction or valence-valence) exchange
and the later is a single exciton or interband (conduction-valence)
effect.\cite{DM,Knox,PB}
The intra-excitonic exchange does not break the symmetry between spin
+1 and spin -1 exciton and has a very weak influence in the exciton
energy levels.\cite{7Sh} In this paper we neglect the intra-ecitonic
exchange in the calculation of the exciton binding-energy. Nevertheless,
the intra-excitonic exchange plays a very important role in the spin
flip mechanism.\cite{7Sh,BAP}

We use a mean field approximation. First, we calculate the expectation
value of the Hamiltonian with a wavefunction equal to the
product of the non-interacting exciton wavefunctions:
\begin{eqnarray}
<H>= \sum _{i} \omega ^{0}_i \ n _{i} + \frac{1}{2} \ \sum _{i,i'} (<i',i| I_d
+ I_{x} | i',i> + <i',i | I_{d} + I_x |i,i'>) \ n_{i'} \ n_{i}
\label{mf}
\end{eqnarray}
with
\begin{eqnarray}
\omega ^{0}_i= <i|T_e + T_h -V_{eh} |i>
\end{eqnarray}
Now, if we make a functional derivation without any further
assumption about the single-exciton
wave function, the Euler-Lagrange equations we obtain are terribly
complicated.\cite{5Ha,6Sc,11Sto} One of the complications is non-locality,
another is our lack of knowledge of $n_{i}$, the quantum non equilibrium
distributions. If we assume ${\bf K}=0$, say, all the excitons are at rest,
nonlocality disappears and instead a set of continuous functions
$n({\bf K},J,M,\nu )$ we have a set of  discrete numbers
$n(J,M,\nu )$. This is the
most drastic assumption we make. In the case of resonant
excitation, the ${\bf K}=0$ hypothesis is more realistic than in the
non-resonant excitation case because, in the resonant regime,
the system receives just the energy required to create the exciton
without any kinetic energy. Experimental information is available both in
the resonant\cite{1LV,2LV,3Aman,4Aman} and in the non resonant\cite{1LV,2LV}
regime. We consider only the resonant case, i.e. ${\bf K}=0$. A less drastic
and more usual assumption is that all the excitons are in their ground state
$\nu =1$. So, spin $M$ is the only quantum number labelling our excitons.
Since holes at the top of the valence band have an angular momentum 3/2 and
electrons a 1/2 one, excitons
have angular momenta running from 2 to -2. Zero angular momentum does
not play any role in the usual experimental configuration of light
propagation along the growth axis in 2D systems.
Hence, in order to describe the system we have only to know four numbers,
i.e., the exciton populations which we denote by the array ${\bf n}
\equiv  \  \{n_{+2},n_{+1},n_{-1},n_{-2} \} \ = \
\{n_{+2},n_{\uparrow},n_{\downarrow},n_{-2} \ \}$.
Another approximation implicit to the derivation of the Euler-Lagrange
equations is that for each {\bf n} the energy takes its equilibrium value.
This is equivalent to say that the changes in {\bf n} are long compared with
collision exciton times: the excitons interact between themselves many
times before the populations change. This approximation is usually
called adiabatic or quasi-equilibrium. Therefore, {\bf n} can be taken
as fixed in the theory to compute energy levels.

Before obtaining the Euler-Lagrange equations, let us discuss the shape
of the Hamiltonian. Substituting {\em i} by $M$ in eq. \ref{mf},
$<H>$ is a sum of integrals, but all the terms
$<I_{d}>$ are zero. This is a nice consequence of the
approximation ${\bf K}=0$ and the neutral charge of the exciton.
We factorize the expectation values $<I_{x}>$ in spin part and
spatial part. Let us first discuss the former and second the later. The spin
part depends on the dimensionality of the space. We shall use the following
notation: $\xi _{s_e,s_h} (M) $ for the probability amplitude of the
electron having spin $s_{e}$ and hole having spin $s_{h}$ in an
exciton with spin M. The spin part of $<I_{x}>$ reads:
\begin{eqnarray}
I^{M_3,M_4}_{M_1,M_2}= \sum _{s_e, s_h,s'_e,s'_h} \xi^{*}_{s_e,s_h}(M_1)
\ \xi^{*}_{s'_e,s'_h}(M_2)\ \xi _{s'_e,s_h}(M_3)\ \xi _{s_e,s'_h}(M_4)
\end{eqnarray}
Obviously the spin wave functions depend on the confinement of the excitons. We
shall make the following approximation for the 2D case
\begin{eqnarray}
\xi_{s_e,s_h}(+2)= \delta_{s_{e} ,\frac{1}{2}} \ \delta _{s_{h},
\frac{3}{2} }\ \ \ \ \
\xi_{s_e,s_h}(+1)= \delta_{s_{e} ,\frac{-1}{2}} \ \delta _{s_{h},
\frac{3}{2} }\nonumber \\
\xi_{s_e,s_h}(-1)= \delta_{s_{e} ,\frac{1}{2}} \ \delta _{s_{h},
\frac{-3}{2} }\ \ \ \
\xi_{s_e,s_h}(-2)= \delta_{s_{e} ,\frac{-1}{2}} \ \delta _{s_{h},
\frac{-3}{2} }
\label{spin}
\end{eqnarray}
For the bulk case $\xi_{s_e,s_h}(M)$ is set equal to the Clebsch-Gordan
coefficient with $J=2$ in analogy with the 2D case in which third
components of the angular momentum equal to $\pm 2$ appear. If the actual
3D exciton wavefunction should be a combination of $J=2$ and $J=1$, some
quantitative changes would appear although the final results would remain
qualitatively unaltered.
Using the eq. \ref{spin}, the only non-zero spin terms for the 2D case are:
\begin{eqnarray}
A_{2D}\equiv I^{1,1}_{1,1}=1 \nonumber
\\ I^{2,1}_{1,2}= I_{1,-2}^{1,-2} = I_{2,1}^{1,2}= I_{-2,1}^{-2,1} =1
\label{2d}
\end{eqnarray}
as well as the ones generated by the following symmetry properties:
\begin{eqnarray}
\begin{array}{c}
I^{\gamma ,\delta }_{\alpha ,\beta }= I^{-\gamma ,-\delta }_{-\alpha ,-\beta
}\\
I^{\gamma ,\delta }_{\alpha ,\beta }=I^{\delta ,\gamma }_{\beta ,\alpha }
\end{array}
\label{sym}
\end{eqnarray}

In the 3D case the only non zero spin terms  are:
\begin{eqnarray}
\begin{array}{lcr}
I_{2,2}^{2,2}=1&A_{3D}\equiv I_{1,1}^{1,1}=
\frac{10}{16}&I_{1,-1}^{1,-1}=\frac{6}{16}\\
I_{2,1}^{2,1}=\frac{3}{4}&I_{2,1}^{1,2}=
\frac{1}{4}&I_{2,-1}^{2,-1}=\frac{1}{4}
\end{array}
\label{3d}
\end{eqnarray}
and the ones generated by eq. \ref{sym}.
The main difference between equations \ref{2d} and \ref{3d} is that
$f_{d}\equiv I_{-1,1}^{-1,1}$ is not zero in the 3D case. These spin term are
proportional to the interaction term $n_{M} n_{-M}$ in eq. \ref{mf}. The
physical consequence of this fact is that, in 2D, M-excitons do not interact
with  $(-M)$-excitons and this enhances  the splitting when  $n(M) \neq n(-M)$.
As it will become clear in section III, if we had $I_{-1,1}^{-1,1}=
I_{1,1}^{1,1}$  then the splitting would be zero.
In order to compare with the experimental results performed in quantum
wells, we shall use our 2D results including a renormalized "$f_d$ term"
($I_{-1,1}^{-1,1}$) as well as a non-interacting exciton binding
energy $E_0^{QW}=E_0^{2D}/2=2E_0^{3D}$.\cite{Bas,Gre}
We take $f_{QW}\simeq6/160$, i. e. one tenth the 3D value.

In order to work out the spatial part, we proceed to make the functional
derivation: $(\delta <H>/\delta \phi_{M} (r))/n(M)=0$
to get four Euler-Lagrange equations, one for each M. In these four
equations there are two well differentiated parts, one  corresponding to
$\delta <H _{0} >/\delta \phi_{M}(r)$ and the other corresponding to
$\delta <H _{int}>/\delta \phi_{M}(r)$. The former generates the
usual Wannier equation \cite{6Sc} and the later generates the
interaction terms. In order to reduce the four interaction terms
$V(e,e'), V(h',h), V(e,h'), V(e',h)$ to one, i.e. V(q),
it is convenient to work in the momentum representation. We are going first to
derive the equations in the case n(2)=n(-2)=0. This will simplify considerably
the equations and will shed some light over the underlying physics.
Besides, in the experiments the $\pm 2$ optically inactive\cite{7Sh}
excitons are less populated than the optically active $\pm 1$. Hence, we have
only two kind of excitons, namely, up and down, $\uparrow $ and $\downarrow $.
We omit the algebra because it does not give any physical information. The
two multiexciton polarized Wannier equations are:(we set $\hbar=1$)
\begin{eqnarray}
\left[ \frac{q^2}{2\mu} - E_{\uparrow}\right] \phi_{\uparrow} (q) - \gamma
_{\uparrow}(q) + \ \ 2A_{d} \ n_{\uparrow} \  \left[ |\phi_{\uparrow}
(q)|^2 \gamma _{\uparrow}(q) - \Sigma_{\uparrow \uparrow}(q)
\phi_{\uparrow} (q) \right] +
\nonumber \\
2f_{d} \ n _{\downarrow} \left[ \gamma _{\downarrow}(q)
\phi_{\downarrow}^{*}(q) \phi_{\uparrow}(q) \ + \
|\phi _{\downarrow} (q)|^2 \gamma _{\uparrow} \ - \Sigma_{\downarrow
\downarrow}(q) \phi _{\uparrow} (q) - \Sigma_{\downarrow \uparrow}(q)
\phi_{\downarrow} (q)\ \ \right] =0
\label{up}
\end{eqnarray}
\begin{eqnarray}
\left[ \frac{q^2}{2 \mu} - E_{\downarrow}\right]\phi_{\downarrow} (q) - \gamma
_{\downarrow}(q) +  \  2A_{d} \ n_{\downarrow} \left[ |\phi_{\downarrow}
(q)|^2 \gamma _{\downarrow}(q) - \Sigma_{\downarrow \downarrow}
\phi_{\downarrow} (q) \right] + \nonumber
\\ 2 f_{d}\ n _{\uparrow} \left[ \gamma _{\uparrow}(q)
\phi_{\uparrow}^{*}(q) \phi_{\downarrow}(q) \ + \
|\phi _{\uparrow} (q)|^2 \gamma _{\downarrow} \ - \Sigma_{\uparrow
\uparrow}(q)
\phi _{\downarrow} (q) - \Sigma_{\uparrow \downarrow}(q) \phi_{\uparrow} (q)\
\ \right] =0
\label{down}
\end{eqnarray}
with
\begin{eqnarray}
\gamma _{s}(q)\equiv \int \frac{d \vec{t}}{(2\pi)^d}\phi_{s}(\vec{t}+\vec{q})
V(\vec{t}) \nonumber \\
\Sigma_{s,s'} (q) \equiv \int \frac{d \vec{t}}{(2 \pi) ^d}
\phi^{*}_{s}(\vec{t} +\vec{q}) \phi_{s'}(\vec{t} +\vec{q}) \ V(\vec{t})
\label{def}
\end{eqnarray}
where $V(\vec{s})$ is the Fourier-transform of the bare Coulomb potential
and $E_{\uparrow }$ and $E_{\downarrow }$ are the Lagrange multipliers
associated to the normalization constraint which are interpreted as the exciton
energies. Note that equations \ref{up}, \ref{down} and \ref{def}
depend on the  dimension through $\phi$, $A_{d}$ and $f_{d}$.
The first two terms in equations \ref{up} and \ref{down} are the usual
Wannier interactionless, two band, exciton equation. The other terms
proportional to $n_{\downarrow}$ and
$n_{\uparrow}$ represent  the mean field  exciton-exciton interaction.
Observe that we can obtain each equation by reversing all the spins of
the other because there is no magnetic field. The spin symmetry breaking
comes from the fact that $n_{M} \neq n_{-M}$.

\section{ Calculation of the interacting exciton energy levels}
\subsection{Perturbation theory}
Solving the multi-exciton Wannier equations \ref{up} and
\ref{down} exactly is not possible and therefore we use some
approximations. First we calculate $E$  perturbatively, using the
the two first terms in equations \ref{up} and \ref{down} as an unperturbed
Hamiltonian and the interacting terms as the perturbation. The
perturbation parameter is obviously $na^d$ and, as we stressed before,
we must drop all the contributions to the enegy with order higher than
$na^d$. Consequently, we do not go further than first order perturbation
theory which means that the interacting exciton wave function is the same
that the non-interacting one. Hence, to first order in perturbation
theory, $\phi_{\downarrow}= \phi_{\uparrow}=\phi_{0}$ where:
\begin{eqnarray}
\phi^{2D}_{0}(q)= \frac{(2 \pi)^{\frac{1}{2}}a}{[1+(aq/2)^2]^{\frac{3}{2}}}
\nonumber \\
\phi^{3D}_{0}(q)= \frac{8(\pi a^{3})^{\frac{1}{2}}}{[1+(aq)^2]^2}
\end{eqnarray}
are the exciton wave functions in the isotropic, parabolic two band
model.\cite{Bass} In the momentum representation the Schr\"{o}dinger
equation is an integral equation.\cite{Cohen} The perturbation
gives a first order correction
\begin{eqnarray}
\Delta E= \int \frac{d \vec{q}}{(2 \pi)^d}
|\phi_0(q)|^2 \ \ \Delta H(q) \ +\int \frac{d
\vec{q}}{(2 \pi)^d}\phi_0(q) \int \frac{d \vec{t}}{(2 \pi)^d} \Delta
H(\vec{t},\vec{p}) \  \phi_0(\vec{t} + \vec{p})
\label{cor}
\end{eqnarray}

Using eq. \ref{up} in eq. \ref{cor} we obtain the following expression
for the 2D case:
\begin{eqnarray}
\Delta E_{\uparrow}= \ 2 n_{\uparrow} \ (I_1 - I_2 ) + 2 \ f_{QW}
\ n_{\downarrow} (I_1 - I_2)
\nonumber \\
\Delta E_{\downarrow} = \ 2 n_{\downarrow} \ (I_1 - I_2 ) + 2 \ f_{QW}
\ n_{\uparrow} (I_1 - I_2)
\nonumber
\end{eqnarray}
where
\begin{eqnarray}
I_1 = \int \frac{d
\vec{q}}{(2 \pi)^d} |\phi_0 (q)| ^{3} \int \frac{d \vec{t}}{(2 \pi)^d}
\ V(\vec{t}) \ \phi_0 (\vec{q} + \vec{t}) \nonumber
\end{eqnarray}
and
\begin{eqnarray}
I_2 = \int \frac{d \vec{q}}{(2 \pi)^d} |\phi_0 (\vec{q})| ^{2} \int
\frac{d \vec{t}}{(2 \pi)^d} | \phi_0 (\vec{q} + \vec{t} )| ^{2} V(\vec{t} )
\end{eqnarray}

Since we are in the lowest order of perturbation theory  $I_{1}$ and
$I_{2}$ do not depend on the perturbed wavefunctions.
In the appendix we show that in 2D $I_{1}= \pi a^2 |E ^{2D} _0|$
and  $I_{2}= \pi |E ^{2D} _0| a^2 \frac{315 \pi ^2}{2^{12}} $,
where $|E ^{2D} _0|$ is the 2D Rydberg
$E^{2D}_0=-2e^2/a=-2h^2/\mu a^2=4E^{3D}_0$. We obtain
\begin{eqnarray}
\Delta E_{\uparrow}= k |E ^{2D} _0| \ a^{2} \ (n_{\uparrow} +
f_{QW} \ n_{\downarrow}) \nonumber \\
\Delta E_{\downarrow}= k |E ^{2D} _0| \ a^{2} \ (n_{\downarrow} +
f_{QW} \ n_{\uparrow}) \nonumber \\
\Delta ^{2D} \equiv \Delta E_{\uparrow} - \Delta E_{\downarrow} =
k |E ^{2D} _0| (1-f_{QW})(n_{\uparrow} - n_{\downarrow} )a^{2}
\end{eqnarray}
where k=3.03. We observe that the inter-excitonic exchange
interaction produces a blue-shift of the levels. Although this calculation
does not include screening effects, $\Delta ^{2D} $ gives properly the spin
splitting because screening is spin independent as discussed below. We observe
that $\Delta ^{2D}$ is proportional to the polarization ($P=(n_{\uparrow}
-n_{\downarrow})/ (n_{\uparrow} +n_{\downarrow}) $) and takes his maximum
value when $f_{QW}=0$, i.e. in the strictly 2D case. Anyway,
$f_{QW}$ is almost negligible compared to 1 and we shall drop it in
the following calculations. In the non polarized case we retrieve
the result obtained by Schmitt-Rink et al. for energy shifts.\cite{SRCM2}

In 3D we obtain analogously $I_{1}-I_{2}=13 \pi/3$ which brings to
\begin{eqnarray}
\Delta E_{\uparrow}= 13 \pi /3 \ |E ^{3D} _0| \ a^{3} \left(\frac{10}{16}
\ n_{\uparrow}+\frac{6}{16} \ n_{\downarrow}\right) \nonumber \\
\Delta E_{\downarrow}= 13 \pi /3 \ |E ^{3D} _0| \ a^{3} \left(\frac{10}{16}
\ n_{\downarrow}+ \frac{6}{16} \ n_{\uparrow}\right) \nonumber \\
\Delta ^{3D} \equiv \Delta E_{\uparrow} - \Delta E_{\downarrow} = 13 \pi
/3|E ^{3D} _0| \left(\frac{10}{16}- \frac{6}{16}\right)(n_{\uparrow}
- n_{\downarrow} )a^{2} =3.4 \ |E ^{3D} _0|(n_{\uparrow} -
n_{\downarrow} )a^{2}
\label{3D}
\end{eqnarray}

The energy splitting between spin $-1$ and spin $+1$ excitons that
occurs in the 3D case is, at equal densities $na^{d}$, very
close to the 2D one both of them scaled in their corresponding
Rydbergs $E_0^d$. Hence, we predict that the energy
splitting, measured in units of $E_0^{QW}$, has a very
weak dependence on the quantum well width. On the other hand, in the
3D non polarized case we do not recover exactly the result obtained
by Haug and Schmitt-Rink \cite{6Sc} for energy shifts because we do not
use spinless wave functions as they do. The effect of the spin part of
the wave function is to set $A_{3D}=10/16$ instead of 1 as in their
calculations.

We have been able to calculate the many-body corrections to the
exciton binding energy in the case $n_{+2}$=$n_{-2}=0$ using
perturbation theory. The many body corrections in $E_{\uparrow}$ and
$E_{\downarrow }$ do not depend on the energy levels of $\pm 2$ excitons.
Therefore, we do not need to write
the multi-exciton equations for the $n_{+2}\neq n_{-2}\neq 0$ case.
In order to avoid tedious algebra we can evaluate the energy
corrections by counting how many integrals are non zero in the sums
of eq. \ref{mf} noticing that each bracket makes a contribution equal to
$I_{1} - I_{2}$ times the spin factor. In this way we obtain for the
2D case the following interexcitonic exchange corrections
\begin{eqnarray}
\delta \left(\begin{array}{c}E_{+2}\\E_{+1}\\E_{-1}\\E_{-2} \end{array}
\right)=
E^{2D}_{0} a^{2}\left(\begin{array}{cccc}
k&k&k&0\\
k&k&0&k\\
k&0&k&k\\
0&k&k&k
\end{array} \right) \left( \begin{array}{c}n_{+2}\\n_{+1}\\n_{-1}\\n_{-2}
\end{array} \right)
\end{eqnarray}
A prediction of this theory is that, neglecting $f_{QW}$-like
factors, the M-excitons interact with equal strenght with all
the others except with the (-M)-excitons for which no interaction exists.
The energy splitting of $\pm 1$ excitons is given by
\begin{equation}
\Delta ^{2D} = k |E^{2D} _0|a^{2} (n_{+1} - n_{-1} )+(n_{+2}-n_{-2}).
\end{equation}
The $\pm2$ exciton density is negligible compared with that of
the $\pm1$ excitons because
the $\pm2$ excitons cannot be optically generated in one
photon processes. Therefore, their influence on $\Delta ^{2D}$ is very small.

In the 3D case we have
\begin{eqnarray}
\delta \left(\begin{array}{c}E_{+2}\\E_{+1}\\E_{-1}\\E_{-2} \end{array}
\right)=
\frac{13\pi}{6} E^{3D}_{0} a^{3}
\left(\begin{array}{cccc}
1&1&\frac{1}{4}&0\\
1&\frac{10}{16}&\frac{6}{16}&\frac{1}{4}\\
\frac{1}{4}&\frac{6}{16}&\frac{10}{16}&1\\
0&\frac{1}{4}&1&1
\end{array} \right) \left( \begin{array}{c}n_{+2}\\n_{+1}\\n_{-1}\\n_{-2}
\end{array} \right)
\label{bul}
\end{eqnarray}
This equation does not carry new physics compared to eq. \ref{3D} and,
as in the 2D case, the influence of the $\pm2$ excitons
is limited because of their very low occupation. In a
situation with $n_{\pm1}\geq 10^{10}cm^{-2}$ and
$\pm2$ exciton negligibly populated, the later energy levels
are higher (less bound) than the $\pm1$ excitons both in 2D
and 3D because the $\pm2$ excitons interact with two higly
populated excitons, say, +1 and -1.

\subsection{ Screenig corrections}
Hitherto we have used the  unscreened Coulomb potential reduced by
the dielectric constant $\epsilon$ of the material in its ground state.
The presence of a considerable amount of excited mobile carriers
(electrons and holes) screens the interacion between these carriers
and the rest of the lattice. The renormalization of the exciton binding
energy caused by screenig has received considerable
attention.\cite{9No,13Kr,Zi76}
In order to simplify the theory we have
calculated the screening corrections in the Random Phase
Approximation (RPA).\cite{6Sc} In this approximation the screening
correction to the binding energy does not depend on the
exciton spins \cite{6Sc,Zi76} being
\begin{eqnarray}
\Delta E^{sc} _{0}  \simeq \ -n \sum_{\nu ',\nu '',q} V^2 (\vec{q}) \
\frac{(|<1s|e^{i\alpha \vec{q} \ \vec{r}} -e^{-i \beta \vec{q} \
\vec{r}} |\nu ''>|)^2\ \ |<1s|e^{-i\alpha \vec{q} \ \vec{r}} -
e^{i \beta \vec{q} \ \vec{r}} |\nu '>|)^2}{2\ (E_{0} + \frac{q^2}{2M_{EX}})}
\end{eqnarray}
where $M_{EX} \equiv m_{e}+m_{h}$ is the total exciton mass and
$|\nu>$ are the exciton internal states, $\alpha\equiv m_{h}/M_{EX}$
and $\beta \equiv m_{e}/M_{EX}$. Using the completeness relation we get:
\begin{eqnarray}
\Delta E^{sc}_{0}  \simeq \ -2n \sum_{\vec{q}} V^2(\vec{q}) \
\frac{(<1s| 1- cos( \vec{q} \ \vec{r}) |1s>)^2}{(E_{0} +
\frac{q^2}{2M_{EX}})} = \nonumber  \\
-2n \ \sum_{\vec{q}} V^2(\vec{q}) \left[1-\frac{1}{(1+a^{2} \
q^{2} /4)^2}\right]^2\ \frac{1}{E_{0} + \frac{q^2}{2M_{EX}}}
\end{eqnarray}
Next, we transform the summation in an integral following the usual
prescription. The screening correction we obtain in the 2D
case is:
\begin{equation}
\Delta E^{sc} _{0}  \simeq \ - \pi \ na^2 |E^{2D}_{0}| w^2 F(w)
\end{equation}
where $w\equiv (M/4\mu)^{\frac{1}{2}}$ and
\begin{equation}
F(w)= \int_{0}^{\infty} \frac{dx}{x} \left[1 -
\frac{1}{(1+x^2)^{3/2}}\right]^2 \frac{1}{x^2 + w^2}
\end{equation}
In a GaAs quantum well $z\equiv \frac{m_h}{m_e} \simeq 2.5$ which
leads to $w^{2}=\frac{(1+z)^2}{4z}\simeq 1.2$, $w^2 F(w)\simeq 0.41$
and $\Delta E^{sc} _{0}  \simeq \ - \ 0.41 \pi \ na^2 |E^{2D}_{0}|$.
The screening contribution reduces the effect of the bare Coulomb
interaction between the carriers producing a relative red-shift of all the
exciton levels and therefore does not contribute to the spin level
splitting. It must be stressed that $w^2F(w)$ is a very smoth funcion
of $m_h/m_e$ and the screening correction is rather insensitive
to variations of this mass ratio.

A physical interpretation of the way the screening modifies the
exciton binding energy is the following. In the previous section we
calculated the exciton binding energies taking into account the
inter-excitonic interaction with the bare Coulomb potential $V_0$.
Now we calculate the dressed Coulomb potential $V_s$ in the RPA and
we treat the difference  $V_{s} - V_{0}$ as a perturbation. It happens
that, to the lowest order, $V_{s} - V_{0}$ is proportional \cite{6Sc}
to $n$ and we must use the non-interacting wave functions to evaluate
the screening correction. Furthermore, we cannot calculate the
screening corrections to the blue-shifts caused by the inter-excitonic
exchange interaction because they would be, at least, of order $(na^2)^2$.
Hence the 2D exciton levels including inter-excitonic
exchange and screening are obtained from
\begin{eqnarray}
\left( \begin{array}{c}E_{+2}\\E_{+1}\\E_{-1}\\E_{-2}\end{array}
\right)= |E^{2D}_{0}|
\left( \begin{array}{c}-1\\-1\\-1\\-1 \end{array} \right)+| E^{2D}_{0}| a^{2}
\left(\begin{array}{cccc}
k-q&k-q&k-q&-q\\
k-q&k-q&-q&k-q\\
k-q&-q&k-q&k-q\\
-q&k-q&k-q&k-q
\end{array} \right)
\left( \begin{array}{c}n_{+2}\\n_{+1}\\n_{-1}\\n_{-2} \end{array} \right)
\label{levels}
\end{eqnarray}
where $k=3.03$ is the bare interexcitonic blue-shift constant and $q=1.28$
is the screening red-shift constant, both dimensionless.

The screening correction energy in bulk was calculated in
RPA by Zimmermann \cite{Zi76} obtaining:
\begin{equation}
\Delta E^{sc} _{0} \simeq \ \pi |E^{3D}_{0}| n a^{3} \left(
\frac{w(32 +63 w +44 w^{2} +11 w^{3}}{(1+w)^4}-
\frac{8w(4+3w)}{(1+w)^2} \right)\equiv \pi |E_{0}| n a^{3} f(w)
\end{equation}
As in the 2D case, the screening correction in the RPA is quite
insensitive to variations of $z$. Hence, we also take $z\equiv
m_h/m_e\simeq2.5$ for 3D GaAs  wich leads $f[w]=-5.0$ and
$\Delta E^{sc} _{0}  \simeq \ -5 \pi |E^{3D}_{0}| n a^{3} $.
Hence, the bulk interacting exciton levels are given by eq. \ref{bul}
minus the screening correction $\Delta E^{sc} _{0}$

\subsection{Variational Approach}

We have a set of complicated equations (\ref{up} and \ref{down})
for which we have applied first order perturbation theory.
We cannot go beyond first order due to
our previous hypothesis but  we would like to extract more
information from those equations. In order to do that we have tried
a simple variational approach in the 2D case.
As we did in the perturbation approach we treat eqs. \ref{up} and \ref{down}
as a Schr\"{o}dinger equation. We identify a Hamiltonian and
minimize $<\phi(q,\alpha) | H |\phi(q,\alpha)>$, with
\begin{equation}
\phi^{var}_{0}(q)= \frac{ (2 \pi)^{\frac{1}{2}}a \ \alpha }
{[ 1+ (a \alpha q / 2 )^2]^{\frac{3}{2}}},
\label{an}
\end{equation}
i. e. we use the exciton radius as a variational parameter.
This Ansatz may be improved if we make $\alpha$ dependent on the
spin. However, the calculations are much simpler with Ansatz \ref{an}.
The non trivial integrals we have to perform are precisely $I_1$ and
$I_2$ scaling $a$ with the variational parameter $\alpha $.
After some algebra we arrive at:
\begin{eqnarray}
\ E_{M}(\alpha ,n, n_{M})= |E^{2D}_{0}| \ \left[\frac{1}{(\alpha)^2}
-\frac{2}{\alpha} + 3.03 \ n_{M} a^{2} \alpha - 1.28 n a^2 (\alpha)^2 \ \right]
\end{eqnarray}
where $n$ is the total density and $n_{M}$ is the M-exciton density.
In this expression we have set $f$=0. Now we have to look for the $\alpha$
that minimizes $E_{M}(\alpha ,n, n_{M})$ for each {\bf n}. We have
done that numerically obtaining that, up to $na^{2}\simeq 0.2$, the
variational technique and the perturbation theory predict the same energy
$\Delta E_(M,{\bf n})$ with an error less than 1$\%$ and $\alpha$
does not differ from 1 (the perturbation theory value) more than
a few percent. We can conclude that,in the small density limit
($na^{2}\leq 0.2$), the energy is properly given by first order
perturbation theory and the wave function is the independent exciton one.

\section{ Exciton spin dynamics}

One could expect our theory to predict new spin flip channels
originated by the interaction. However, this is not the case
because the interaction terms in eqs. \ref{up} and \ref{down} are
proportional to $n$ and the transitions rates are proportional to
the squared interactions terms, i.e. $n^{2}$. Hence, following the
considerations made in section II, we {\em must} neglect these
'interacting' transition rates: our theory predicts no significant
variations of transitions rates with respect to those of the
non-interacting theory \cite{7Sh} while the energy levels are
the corresponding to equation \ref{levels}. Therefore, we borrow
the population evolution equation from ref. \cite{7Sh}
\begin{equation}
\frac{d {\bf n}}{dt}= {\cal W} \ {\bf n}
\label{pop}
\end{equation}
where
\begin{eqnarray}
{\cal W} \equiv \left( \begin{array}{cccc}
-(W^{e}_{2,1}+W^{h}_{2,-1})&W^{e}_{1,2}&W^{h}_{-1,2}&0\\ %
W^{e}_{2,1} & \begin{array}{c}-(W_{R} +W^{1,-1}_{ex} \\
+W^{e} _{1,2} +W^{h}_{1,-2}) \end{array}  &W^{-1,1}_{ex}& W^{h}_{-2,1}\\ %
W^{h}_{2,-1}&W^{1,-1}_{ex}& \begin{array}{c}-(W_{R}+W^{-1,1}_{ex} \\
+W^{e}_{-1,-2}+W^{h} _{-1,2}) \end{array}&W^{e}_{-2,-1}\\ %
0&W^{h}_{1,-2}&W^{e}_{-1,-2}&-(W^{e}_{-2,-2} + W^{h}_{-2,-1}) \end{array}
\right).
\end{eqnarray}
This equation is obtained from the master equation by making the
approximation of quasi-equilibrium and taking into account the
detailed balance principle.\cite{Van} There are three kinds of
transition rates: the easiest to understand is the radiative
recombination rates $W_{R}$ which affect only to the optically
active excitons $\pm1$. Also, we have the $W^{e(h)}_{M,M'}$
rates, say, the transition (M)-exciton to (M')-exciton caused by
the {\em e(h)} spin flip. The last type of transition rate,
$W^{M,M'}_{ex}$ is associated to the intra-excitonic exchange
mechanism. Following ref. \cite{7Sh} we  set
\begin{eqnarray}
W_{R}=1/400 ps \nonumber \\
W^{e(h)}_{M,M'}=\frac{1}{\tau^{e(h)}} \ \frac{1}{1+exp[(E_{M'}-E_{M})\beta]}
\nonumber \\
W^{M,M'}_{ex}=\frac{1}{\tau^{ex}} \ \frac{1}{1+exp[(E_{M'}-E_{M})\beta]}
\end{eqnarray}
where $\tau^{e,h}$ are the single particle spin flip times,\cite{7Sh,BAP,DP}
$\tau^{ex}$ is the exchange spin flip time
calculated by Maialle et al.,\cite{7Sh} $\beta=1/K_{B}T$
and $E_{M}$ are those of eq. \ref{levels}. The numerical values of
$\tau^{e(h)}$
and $\tau^{ex}$ are taken from the case II of the reference.\cite{7Sh}

\section{Results and Conclusions}

The solution of the non-linear equations \ref{pop} is obtained
numerically by a Runge-Kutta method. Inserting the
time dependence of the densities in eq. \ref{levels} we obtain the time
evolution of the exciton levels. In our theoretical calculations
the natural energy scale is the Rydberg (the 2D or 3D ).
In order to compare our results with experiments
\cite{1LV,2LV,3Aman,4Aman,Stark} we shall plot the energies
in units of $E_0^{QW}$. On the other hand, we have
checked that the results are insensitive to the variation of the
initial  populations of $\pm2$-excitons provided they are less than
$10\% $ of the total initial density , $n_{0}$ and, consequently,
we present figures obtained from an initial density of $\pm 2$-excitons
equal to zero. We adopt the following  conventions:
i) the $(+1)$-exciton is the more populated state at $t=0$; ii) the
populations are measured in $n_{0}$ units; iii) the origin of energies is
taken at the bottom of the conduction band.

In figure \ref{fig1} we plot our theoretical predictions of populations
(a) and energy levels (b) with an initial density of
$na^{2}=0.1$ (about $10 ^{11}$ excitons per $cm^2$) and a initial
polarization $P=(n_{\uparrow} -n_{\downarrow})/(n_{\uparrow}
+n_{\downarrow}) =80 \% $. In figure \ref{fig1}(a) we observe that the
polarization disappears in roughly 50ps. As we see in the
inset the $\pm2$ populations are, at least, one order of magnitude
smaller and they decay much slower than the optically
active ones. In figure \ref{fig1}(b) the more important features
are the spliting between spin $+1$ and spin $-1$ excitons and the
fact that the $\pm2$-excitons levels are closer to the conduction
band than the $\pm1$ ones. In t=0 the splitting takes its maximum value,
$0.45E_0^{QW}$, about 3 meV, and then
decreases becoming zero at t=50 ps. It must be stressed that the
copolarized (more populated) exciton level shows an energy blue-shift and
the contrapolarized (less populated), when is very low populated,
starts having a red-shift and, when its population increases, presents
a blue-shift as observed experimentally.\cite{4Aman} We also show
similar results with different either initial polarization
(Fig. \ref{fig2}) or population (Fig. \ref{fig3}).
The main difference between Fig. \ref{fig2} where $P=60 \% $ and
Fig. \ref{fig1} where $P=80 \%$ is the decrease of the $\pm1$
splitting as well as a shorter duration of this splitting.
In figure \ref{fig3} we set the initial population
$na^{2}$ equal to $0.05$ (about $5.0 \ 10 ^{10}$ excitons per
$cm^2$) and the initial polarization once again $P=80\%$. The splitting is
smaller than the one observed in the Fig. \ref{fig1} , about $0.25E_0^{QW}$
and its duration is longer.

 The main consequences we extract from these results are the following:

i) the spin level degeneracy breaking is proportional to the
polarization of excitons populations and both polarization and
splitting disappear in a time of the order of 50 ps

ii) the $\pm2$ excitons are negligibly populated and therefore do
not influence the $\pm1$ exciton energy levels although the opposite
is true: the $\pm1$ excitons do influence the $\pm2$ energy levels.

In order to compare with experimental information, an interesting
way of presenting our results is to give
energy levels at fixed times as a function of the initial exciton
density. In figure \ref{fig4} some of the conclusions
we stated before become more clear. The splitting increases with $P$ and $n$.
We also observe how the red-shift of the spin $-1$ exciton at an early
stage ($10ps$) becomes a blue-shift at $25ps$.
Those splittings have been observed experimentally\cite{1LV,2LV} but
there is significant difference with our results. Some experiments\cite{munoz}
seem to suggest that
the majority carriers level remains essentially constant as a function
of the initial density, something that clearly does not happens in
our results. Moreover, our theory gives good results for level splittings but
fails to describe the absolute position of the excitonic luminescence peaks
as a function of the initial carrier density.\cite{1LV,2LV} We believe
that these disagreements can be due to the pinning of excitons
to impurities, an effect that might change the exciton energies and
that has not been taken into account in our theory. Another
possible explanation for the mentioned disagreements could be the failure of
the approximation of considering the center of mass at rest to explain
non-resonant excitation experiments.

In summary, we have presented a theory which describes the interacting
polarized exciton gas with moderately high density. We have obtained a
set of multi-exciton Wannier equations that we have solved, both
perturbativately and variationally, obtaining blue-shifts originated by the
inter-excitonic exchange interaction. Whenever differences between the exciton
populations exist, energy levels are splitted. Calculated
and experimentally observed splittings are in good qualitatively agreement.

\section {Acknowledgments}

This reserch was supported in part by the Comisi\'on Interministerial
de Ciencia y Tecnolog\'{\i}a of Spain under contract MAT 94-0982-C02-01
and by the Commission of European Communities under contract
Ultrafast CHRX-CT93-0133.

\appendix
\section{\bf Calculation of $I_{1}$ and $I_{2}$ }

In this appendix we calculate $I_1$ and $I_2$ both in 2D and 3D.
The calculation of
\begin{eqnarray}
I_1 = \int \frac{d
\vec{q}}{(2 \pi)^d} |\phi_0 (q)| ^{3} \int \frac{d \vec{t}}{(2 \pi)^d}
\ V(\vec{t}) \ \phi_0 (\vec{q} + \vec{t})
\label{A1}
\end{eqnarray}
can be done both in 2D and 3D by using the Schr\"odinger equation
\begin{eqnarray}
\int \frac{d \vec{t}}{(2 \pi)^d}
\ V(\vec{t}) \ \phi_0 (\vec{q} + \vec{t}) = \left[ \frac{q^2}{2\mu} -
E_{0}\right] \phi_{0} (q)
\label{A2}
\end{eqnarray}
Substituting \ref{A2} in \ref{A1} we obtain
\begin{eqnarray}
I_1= \int \frac{d
\vec{q}}{(2 \pi)^d} |\phi_0 (q)| ^{4} \ \left[ \frac{q^2}{2\mu} - E_{0}\right]
\end{eqnarray}
In the 2D case we obtain for this integral
\begin{eqnarray}
I_1= 8 \pi a^2 |E^{2D}_0| \int_{0}^{\infty}\frac{x \ dx}{(1+x^2)^5}
= \pi |E^{2D}_0| a^2
\end{eqnarray}
and in the 3D case:
\begin{eqnarray}
I_1= 2^{11} |E^{3D}_0| a^3 \int_{0}^{\infty}\frac{x^2 dx}{(1+x^2)^7}
=21 \pi |E^{3D}_0| a^3.
\end{eqnarray}
The calculation of
\begin{eqnarray}
I_2=\int \frac{d \vec{q}}{(2 \pi)^d} |\phi_0 (\vec{q})| ^{2}
\int \frac{d \vec{t}}{(2 \pi)^d} | \phi_0 (\vec{q} + \vec{t})| ^{2}V(\vec{t})
\label{I2}
\end{eqnarray}
is more cumbersome. We start with the 2D case by
perfoming the integral
\begin{eqnarray}
\Sigma_{00}\equiv\int \frac{d \vec{t}}{(2 \pi)^2} | \phi_0 (\vec{q} +
\vec{t})| ^{2}V(\vec{t})=\int\frac{d \vec{s}}{(2 \pi)^2} V(\vec{s} )
\int d\vec{r} \chi(\vec{r} ) e^{-i(\vec{s}+\vec{q}) \vec{r} } = \nonumber \\
\int d\vec{r}  \chi (\vec{r}) e^{-i\vec{q}\vec{r}} \int
\frac{d\vec{s}}{(2 \pi)^2} V(\vec{s}) e^{-i\vec{s}\vec{r}}
\label{sig}
\end{eqnarray}
where
\begin{eqnarray}
\chi(\vec{r}) \equiv \int\frac{d \vec{t}}{(2 \pi)^2} |
\phi_0 ( \vec{t} )|^2  e^{i \vec{t} \vec{r}} = \nonumber \\
\frac{1}{2\pi} \int_{0}^{\infty} t J_{0}(tr)
|\phi_0 (t)|^{2} = a^2 \int_{0}^{\infty} \frac{t J_{0}(tr) dt }
{[1+ a t/2)^2]^3}= \frac{\gamma^2}{2} K_{2}(\gamma)
\label{chi}
\end{eqnarray}
where $J_{0}(x)$ is the Bessel function of the first kind,
$K_{2}(\gamma)$ is the Modified Bessel function and $\gamma \equiv
2r/a$. Since $V(\vec{s})$ is the Fourier-transform the Coulomb potential,
\begin{eqnarray}
\int \frac{d\vec{s}}{(2 \pi)^2} V(\vec{s}) e^{-i\vec{s}\vec{r}}
= \frac{e^2}{r}
\label{coul}
\end{eqnarray}
and we obtain
\begin{eqnarray}
\Sigma_{00}= \frac{4 \pi e^2}{a^2} \int_{0}^{\infty} r^2 K_{2}
(\frac{2r}{a}) J_{0}(qr) dr=
3 \pi^2 \ e^2 a \  _{2}F_{1}[ \frac{1}{2}, \frac{5}{2},1,- (qa/2)^2]
\label{sigex}
\end{eqnarray}
where
\begin{eqnarray}
_{2}F_{1}[a,b,c,x]\equiv \sum_{k=0}^{\infty} \frac{(a)_k (b)_k }{(c)_k }
\frac{x^k}{k!} ; (a_k)\equiv \frac{(a+k-1)!}{(a-1)!}
\end{eqnarray}
is the confluent Hypergeometrical function. In order to
obtain $I_{2}$ we use \ref{sigex} in \ref{I2} and we get
\begin{eqnarray}
I_{2}= 3 \pi^2 \ a \ e^2\int_{0}^{\infty} \frac{x \
_{2}F_{1}[ \frac{1}{2}, \frac{5}{2},1,- x^2] \ dx }{(1+x^2)^3}=
\frac{315}{4096} \pi^3 |E_0^{2D}| a^2
\end{eqnarray}
where we have used
\begin{eqnarray}
\int_{0}^{\infty} \frac{x \  _{2}F_{1}[ \frac{1}{2}, \frac{5}{2},1,-
x^2] \ dx }{(1+x^2)^3}=
\frac{2}{3} \pi \frac{315}{4096}
\end{eqnarray}
In 3D the calculation of $I_{2}$ is easier. We have
\begin{eqnarray}
I_2= \int \frac{d \vec{q}}{(2 \pi)^d} |\phi_0 (\vec{q})| ^{2}
\int \frac{d \vec{t}}{(2 \pi)^d} | \phi_0 (\vec{q} + \vec{t} )| ^{2}
V(\vec{t} )\nonumber\\
=\frac{2^{11}}{\pi} e^2 a^2 \int_{0}^{\infty} \frac{x^2 dx}
{(1+x^2)^4} \int_{-1}^{+1} d\xi \int_{0}^{\infty} \frac{dy}
{\left(1+x^2 +y^2 - 2 xy \xi\right)^4}
\end{eqnarray}
where $x\equiv a|\vec{q}|$, $y\equiv a |\vec{t}|$ and $a^2
\vec{q}\ \vec{t}= xy cos(\xi)$.
Now we perform the integrations over $\xi$ and $y$ in two steps
\begin{eqnarray}
F(x,y) \equiv \int_{-1}^{+1}\frac{d\xi}{\left(1+x^2+y^2-2xy \xi
\right)^4}=  \nonumber \\
\frac{1}{6xy} \left\{ \frac{1}{\left(1+x^2 +y^2 -2xy \right)^3}-
\frac{1}{\left(1+x^2 +y^2 +2xy \right)^3}\right\}
\end{eqnarray}
and
\begin{equation}
\int_{0}^{\infty} \frac{dy}{y} F(x,y)= \frac{\pi}{8}x
\frac{15+10x^2 +3 x^4}{(1+x^2)^3}\equiv J(x).
\end{equation}
Hence, we obtain
\begin{eqnarray}
I_{2}=\frac{2^{12}}{6\pi} |E_{0}^{3D}| a^3 \int_{0}^{\infty}
\frac{x \ J(x)dx}{\left[1+x^2 \right]^4}=\frac{100}{6} \pi \ |E_{0}^{3D}|a^3
\end{eqnarray}

\begin{figure}
\caption{ (a) $\pm1$ exciton population as a function of time (0-400 ps).
The dashed line is the $-1$ population. In the inset we plot the +2 (circles)
and -2 (crosses) populations (notice the different time scale).
(b) Exciton levels as a function of time (0-140 ps) measured in quantum
well free exciton energy (see text). Crosses and circles are again the
$\pm2$ excitons. The dashed line correspond to (-1)-exciton energy.
Initial density $na^{2}=0.1$, initial polarization P=$80 \%$.}
\label{fig1}
\end{figure}

\begin{figure}
\caption{As in fig. 1 with n=0.1, $P=60 \% $ }
\label{fig2}
\end{figure}

\begin{figure}
\caption{As in fig. 1 with n=0.05, $P=80\% $}
\label{fig3}
\end{figure}

\begin{figure}
\caption{$\pm 1$ exciton energies at (a) $t=10ps$ and (b)
$t=25ps$ as a function of
initial densities. The solid and dashed lines correspond to an
initial polarization of $80\% $ and $60\% $ respectively, and
upper and lower lines correspond to spin $+1$ and  spin $-1$ exciton
respectively.}
\label{fig4}
\end{figure}



\begin{references}
\bibitem{coms} Following the habitude in the field, we denote the third
component of the
exciton angular moment with  the word {\em spin}.
\end{references}
\end{document}